\documentclass[sigconf]{acmart}
\AtBeginDocument{%
  }

\usepackage{usebib}
\usepackage{graphicx}
\usepackage{multirow}
\usepackage{makecell}
\usepackage{caption}
\usepackage{subcaption}
\usepackage[inline]{enumitem}

\setcopyright{iw3c2w3g}
\acmYear{}
\acmDOI{}
\acmConference[]{}{}{}
\acmISBN{}
\begin{document}

\title{Solving cold start in news recommendations: a RippleNet-based system for large scale media outlet}

\author{Karol Radziszewski}
\email{karol.radziszewski.dokt@pw.edu.pl}
\orcid{0009-0006-9408-0508}
\authornote{Corresponding author}
\affiliation{
  \institution{Warsaw University of Technology}
  \city{Warsaw}
  \country{Poland}
}
\affiliation{
  \institution{Ringier Axel Springer Tech}
  \city{Warsaw}
  \country{Poland}
}

\author{Micha\l{} Szpunar}
\email{michal.szpunar@ringieraxelspringer.pl}
\affiliation{
  \institution{Ringier Axel Springer Tech}
  \city{Warsaw}
  \country{Poland}
}

\author{Piotr Ociepka}
\email{piotr.ociepka@ringieraxelspringer.pl}
\orcid{0009-0004-9242-8887}
\affiliation{
  \institution{Ringier Axel Springer Tech}
  \city{Krakow}
  \country{Poland}
}
\affiliation{
  \institution{AGH University}
  \city{Krakow}
  \country{Poland}
}

\author{Mateusz Buczy\'nski}
\email{mp.buczynski2@uw.edu.pl}
\orcid{0000-0003-1734-9287}
\affiliation{
  \institution{Ringier Axel Springer Tech}
  \city{Warsaw}
  \country{Poland}
}
\affiliation{
  \institution{University of Warsaw}
  \city{Warsaw}
  \country{Poland}
}
\renewcommand{\shortauthors}{Radziszewski et al.}

\begin{abstract}
We present a scalable recommender system implementation based on RippleNet, tailored for the media domain with a production deployment in Onet.pl, one of Poland’s largest online media platforms. Our solution addresses the cold-start problem for newly published content by integrating content-based item embeddings into the knowledge propagation mechanism of RippleNet, enabling effective scoring of previously unseen items. The system architecture leverages Amazon SageMaker for distributed training and inference, and Apache Airflow for orchestrating data pipelines and model retraining workflows. To ensure high-quality training data, we constructed a comprehensive golden dataset consisting of user and item features and a separate interaction table, all enabling flexible extensions and integration of new signals. 
\end{abstract}





\maketitle

\section{Introduction}

The rapid pace of content publication in news and media platforms introduces significant challenges for recommender systems, particularly due to the continuous emergence of new items and sparse user interactions. 
These challenges, commonly known as the cold-start problem, limit the effectiveness of traditional approaches that rely heavily on historical interaction data. 
To bridge this gap, we propose an enhanced recommendation approach leveraging RippleNet, a knowledge-aware recommender framework. 
Our solution integrates RippleNet's entity embeddings with text embeddings generated by large language models (LLMs), enabling effective semantic representation and recommendation of newly introduced content, hence addressing the cold-start limitations visible in highly dynamic domains such as news recommendation.

\section{Literature review}


\subsection{Cold start Problem}

A fundamental challenge in recommendation is the cold start problem, which occurs when no observable data about the user or an item is available in the past \cite{DePessemier2025}. 
On the user's end, a practical fallback is to recommend generally popular or trending items until more personalized signals are collected. 
In new item cold-start cases, a freshly added item has no interactions yet, most commonly content-based techniques are used, since item attributes or content can serve as latent space, which “already solves a major part” of the problem \cite{KARIMI20181203}. 
Metadata-driven approaches (using item descriptors like genre, tags, or knowledge from knowledge graphs) and hybrid models (which combine content-based scores with collaborative signals) are widely used to give new items an initial score.

Another angle is to use the exploration-exploitation tradeoff: for instance, deploying a bandit algorithm that occasionally promotes new items to gather feedback. 
However, when only implicit feedback (clicks, impressions, etc.) is available, there are no explicit negative signals, making it harder to discern a new user’s dislikes. 
In highly dynamic domains like news and media streaming, new items arrive continuously and have short lifespans.
This creates a permanent item cold-start scenario \cite{saveski2014item}. 
Conventional collaborative filtering models (which assume a relatively static item catalog) struggle here \cite{Li_2010}. 
Research has addressed this by frequent model updates and by prioritizing recent information. 

\subsection{Knowledge-Aware Recommendations}

In recent years, knowledge-aware recommender systems have gained attention as a solution to data sparsity and cold-start issues. The key idea is to incorporate side information in the form of knowledge graphs (KG) about items and users. Traditional approaches fell into two categories: 
\begin{enumerate*}
    \item embedding-based methods, which learn latent representations of entities (items and their related attributes);
    \item path-based methods, which explicitly search for connectivity paths between users and items via intermediate knowledge nodes.
\end{enumerate*}
While effective to a degree, each has limitations: embedding-based models may not capture complex relational semantics, and path-based models can be ad-hoc or hard to scale.

\begin{figure}[h!]
    \centering
    \caption{RippleNet's ripple sets in a knowledge graph from exemplary dataset}
    
    \includegraphics[width=0.4\linewidth]{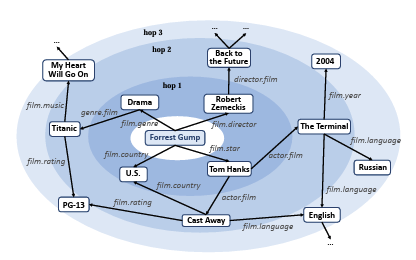}

    \textbf{Source:} \textit{\citefullauthor{ripple}, \usebibentry{ripple}{title}, \usebibentry{ripple}{year}}

    \label{fig:ripplenet}
\end{figure}

RippleNet \cite{ripple} emerged as an influential knowledge-aware framework designed to overcome these limitations. 
It treats the knowledge graph as a base of interconnected entities and relations (e.g. items, their attributes, categories, actors, etc.). 
The model introduces a "ripple" propagation mechanism inspired by the ripple effect on water. 
Given a user’s known interests, RippleNet activates a set of entities in the knowledge graph that are directly linked to those interests (first-hop neighbors). These might be, for example, the authors, topics, or brands associated with the user’s consumed items. Then, in an iterative fashion, the model expands further (second hop, third hop, etc.), activating entities connected to the first set, and so on (this process is graphically visualized in the figure \ref{fig:ripplenet}). Each “ripple” thus spreads the user’s preference signal further through the graph. By the end of $H$ hops, multiple waves of activated entities – triggered by different historical items – collectively form a rich preference distribution over the knowledge graph. For a given candidate item to recommend, RippleNet estimates the user’s preference by how strongly these ripples intersect with that item’s knowledge profile. This end-to-end architecture showed substantial accuracy gains across several domains, including movies, books and even news recommendation, outperforming earlier baselines. 

\bibinput{sample-base}

Knowledge-aware models are especially beneficial in cold-start situations; for example, even if a new item has no users yet, its connections in a knowledge graph (such as its category, creator, or related items) can be used to find potential interested users. Likewise, a new user’s few initial clicks can be mapped to knowledge entities (topics, etc.) to quickly broaden the understanding of their interests. They have also been applied in news recommendation. For example, \cite{wang2018dkndeepknowledgeawarenetwork} proposed DKN: Deep Knowledge-Aware Network. It combines news text understanding with entity information from a knowledge graph. It employs a convolutional neural network  where each news article is represented by two aligned sequences: words (in the title/body) and entity mentions (mapped to a knowledge graph embedding) - the model learns a fused semantic-knowledge representation of news content. An attention module then aggregates a user’s reading history in a personalized way, emphasizing those past articles most relevant to the candidate news. 

\subsection{Recommender Systems for the News Domain}

Recommender systems in news and media face distinct challenges due to the highly dynamic nature of the content, characterized by continuous publication and rapid obsolescence of news articles. Unlike static domains such as movies or books, this domain perpetually operates in a state of item cold-start, demanding strategies that emphasize recent and trending content. Classic collaborative filtering approaches struggle under these conditions due to sparse interaction data and the necessity for rapid model updates. To manage this, leading platforms such as Google News \cite{10.1145/1242572.1242610} developed scalable, hybrid approaches combining clustering techniques, latent semantic models, and real-time co-visitation metrics to continuously adapt recommendations to user interactions and item freshness.

Addressing these challenges, our approach employs RippleNet to capture deeper semantic relationships among items. Specifically, we enhance RippleNet by mapping embeddings learned during model training directly onto text embeddings generated by large language models (LLMs). This allows for effective representation of new items, significantly mitigating cold-start issues by capturing inherent semantic content similarities even without user interaction history.

\section{Data}
We constructed "golden" RASP recommendation dataset in the unified Recbole \cite{recbole} unified format, comprising three core files:
\begin{itemize}
    \item \texttt{.inter}, which logs user–item interactions and interaction-related features; 
    \item \texttt{.user}, detailing user attributes;
    \item \texttt{.item}, encompassing item metadata.
\end{itemize}
To support scalability research, three dataset versions - small, medium, and large - are provided, with their descriptive statistics summarized in Table \ref{tab:rasp_dataset_stats}. 
Notably, this dataset is constructed from real-world interactions in the Polish news service and Polish language, a Slavic language family member, unlike many existing datasets based on English. This linguistic distinction introduces unique semantic, morphological, and lexical challenges relevant to natural language processing and recommendation modeling, particularly in content-based and embedding-based approaches. We added Stylometrix-based features \cite{okulska2023stylometrixopensourcemultilingualtool} to catch this linguistic domain. The dataset spans from March 12, 2025, to April 14, 2025, and is split into training, validation (the three days preceding the test period), and testing (the final three days) subsets.

The dataset’s \texttt{.inter} file contains interaction-level data, including:
\begin{enumerate*}[label=(\roman*)]
  \item \texttt{user\_id} and \texttt{item\_id} identifiers,
  \item \texttt{is\_click}, the binary recommendation target (1 for click, 0 for impression only),
  \item temporal features such as \texttt{event\_timestamp\_unix}, \texttt{event\_date}, \\ \texttt{weekday}, \texttt{hour}, and \texttt{is\_business\_day} (based on Polish holidays),
  \item engagement metrics like \texttt{ip\_cnt}, \texttt{pu\_ip\_cnt}, \texttt{pv\_cnt}, \texttt{glowna\_ ip\_cnt}, \texttt{pv\_on\_content\_publication\_premium\_cnt}, and \texttt{ses\_ \\duration\_sum},
  \item device and session features such as \texttt{device\_type}, \texttt{context\_client\_brand}, \texttt{context\_client\_version}, \texttt{cs} (screen size), and \texttt{rh\_user\_agent},
  \item geolocation approximations via \\ \texttt{latitude}, \texttt{longitude}, \texttt{accuracy\_radius}, and regional identifiers (\texttt{geoip\_city\_name},  \texttt{geoip\_region\_name}),
  \item user state flags, e.g., \texttt{is\_active\_click}, \texttt{is\_active\_pageview}, \texttt{user\_evoked\_sso\_logged\_ in}, and \texttt{user\_ subscriber},
  \item a predictive feature like \texttt{lts\_pred} (likelihood to subscribe to Onet Premium).
\end{enumerate*}

The \texttt{.user} file user profile information, such as:
\begin{enumerate*}[label=(\roman*)]
  \item user BERT-based reading preferences for 10 categories calculated as precentage of read articles within a category,
  \item a login status (\texttt{user\_sso\_name}),
  \item user browser and device.
\end{enumerate*}

The \texttt{.item} file captures rich article-level content and metadata, including:
\begin{enumerate*}[label=(\roman*)]
  \item text-based fields: \texttt{title}, \texttt{lead}, \texttt{text}, and derived metrics like \texttt{text\_length},
  \item authorship and multimedia: \texttt{author}, \texttt{images},
  \item semantic annotations via \texttt{wikidata\_entities\_words}, \texttt{wikidata\_ entities\_ids},  \texttt{wikidata\_entities\_scores}, \texttt{wikidata \_topics}, and \texttt{wikidata\_topics\_scores},
  \item premium content status: \texttt{content\_publication\_premium},
  \item neural embeddings (\texttt{openai \_embedding}, 256-dimensional, computed via OpenAI’s \\ \textit{text-embedding-3-large}),
  \item internal BERT-based category confidence scores for 10 categories,
  \item stylometrix features for \texttt{title}, \texttt{lead}, and \texttt{text}, capturing linguistic, syntactic, and lexical complexity of Polish language.
\end{enumerate*}

The “golden” label reflects the dataset’s high quality and rigorous preprocessing. By aligning with Recbole’s format, the dataset enables immediate compatibility with numerous state-of-the-art recommendation models and tools, fostering reproducibility and streamlined benchmarking within RASP's recommendation system.


\begin{table}[ht!]
\centering
\footnotesize
\caption{Dataset statistics for RASP Recommendation Data, including interaction labels. Last two columns shows number of positive (clicks) and negative (recommendations without click) interactions.}
\begin{tabular}{llrrrr}
\toprule
\textbf{Dataset} & \textbf{Split} & \textbf{Users} & \textbf{Items} & \textbf{Pos (1)} & \textbf{Neg (0)} \\
\midrule
Small    & Train & 18,727 & 20,794 & 498,973 &  4,823,570 \\
         & Valid & 8,033  & 3,653  & 62,496 &  655,422 \\
         & Test  & 7,560  & 2,845  & 52,160 &  456,302 \\
\addlinespace
Medium   & Train & 74,743 & 20,800 & 1,975,109 & 18,814,474 \\
         & Valid & 32,393 & 3,653  & 247,652 &  2,616,881 \\
         & Test  & 29,941 & 2,846 & 208,294 & 1,815,687 \\
\addlinespace
Big      & Train & 298,526 & 20,800 & 7,900,163 & 75,702,502 \\
         & Valid & 129,146 & 3,653  & 988,923 & 10,411,131 \\
         & Test  & 119,654 & 2,846  & 827,126 & 7,233,277 \\
\bottomrule
\end{tabular}
\label{tab:rasp_dataset_stats}
\end{table}

\section{Cold-start solution}

The biggest challenge in addressing the cold start problem in the RippleNet model used for news recommendations is its inability to recommend entities (i. e., articles) which were unknown or non-existent during the model training. To overcome this limitation, we developed a method for translating semantic information about a new article into the representation space of the RippleNet model. Since the RippleNet model represents each entity as an internal embedding and our knowledge about semantics of each article is based on embeddings generated by Large Language Models (LLMs) --- such as OpenAI \cite{openai_embeddings} models or domain-specific alternatives like PolBERT \cite{Kleczek2020} --- we decided to verify two distinct approaches, one based on neural encoder and second based on replacing unknown embeddings with the closest known ones.

\subsection{Neural encoder}
The first approach was to train a neural encoder to predict the RippleNet embedding from the corresponding LLM-based embedding. The encoder was trained as a fully connected feedforward network with two hidden layers. We evaluated two loss functions, mean squared error (MSE) and cosine similarity. As the latter yielded better results in terms of alignment with the RippleNet space, we decided to ditch MSE. The training and evaluation datasets consisted of pairs of both RippleNet and corresponding LLM embeddings of articles present in the RippleNet training set. Exemplary results are in figure \ref{fig:embedding_sim}. It's a histogram of cosine similarities between trained embeddings and encoded LLM ones.

\subsection{Similarity-based replacement}
\begin{figure}
\caption{Histogram of cosine similarity between matched and real embedding}
\centering
\begin{subfigure}{.25\textwidth}
  \centering
  \includegraphics[width=.4\linewidth]{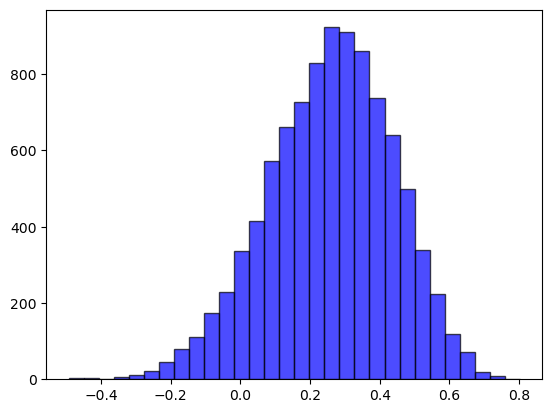}  
  \caption{using neural encoder}
  \label{fig:sub1}
\end{subfigure}%
\begin{subfigure}{.25\textwidth}
  \centering
  \includegraphics[width=.4\linewidth]{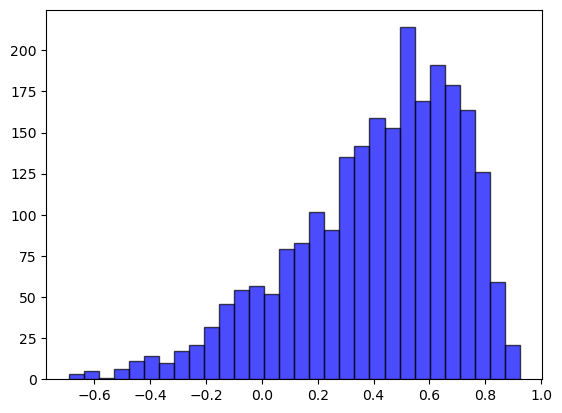}
  \caption{using cosine similarity matching}
  \label{fig:sub2}
\end{subfigure}

\label{fig:embedding_sim}
\end{figure}

In contrast to the neural encoder approach, the similarity-based replacement method avoids the use of artificial non-existent embeddings as input to the RippleNet model. Instead, it computes the cosine similarity between the LLM-based embedding of the new item and the LLM-based embeddings of all known items. The RippleNet embedding corresponding to the most similar known item is used as input to the recommender model. Exemplary results are in figure \ref{fig:embedding_sim}. We notice a larger number of instances close to real embedding, hence we chose this setting for further testing.

\section{Methodology}
\subsection{Offline evaluation}
When evaluating offline the similarity-based recommendation solution - which is better suited for deployment in large-scale online environments - we use our production model as a baseline. The production system employs a deep neural network for click prediction and is described in our prior work \cite{RadziszewskiOciepka2024}. Since RippleNet requires retraining daily to remain effective in dynamic domains such as news recommendation, we performed an offline evaluation by training it on a single day and testing it across three temporal slices: the day before training, the training day itself and the day after.
\subsection{Online evaluation}
\begin{figure}[ht!]
    \caption{The pipeline illustrates the automated training and deployment process for the RippleNet model.}
    \centering
    \includegraphics[width=0.66\linewidth]{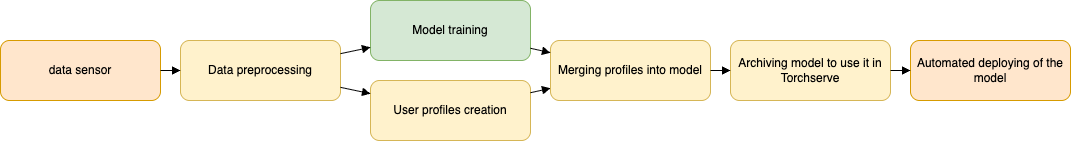}
    \footnotesize
    
    Orange nodes represent Airflow tasks, yellow nodes correspond to SageMaker processing jobs, and the green node denotes the SageMaker training job.

    \label{fig:airflow-dag}
\end{figure}

The news recommendation environment evolves rapidly, necessitating daily retraining of the RippleNet model to continuously integrate new user-item interactions and enable its evaluation in an online setting. We orchestrate the retraining process using an Airflow DAG, which automates both data processing and model training tasks executed on AWS SageMaker. The overall workflow is depicted in Figure \ref{fig:airflow-dag}. The pipeline begins with a data availability check, which ensures that data is present. Once the prerequisite data is available, next task extracts knowledge graph and reformat data to match Recbole format. 
The preprocessed data is then consumed by two parallel tasks: training, which performs model training on GPU using the sample of latest data, and user profiles creation, which constructs ripple sets at the scale of millions of users. Their outputs are aggregated in the merge task. Next, the torch-model-archiver module packages the trained model into a format suitable for deployment. Finally, the deployment task, deploys the archived model to the production environment, enabling real-time inference and online evaluation.


\section{Results}
\subsection{Offline evaluation}
The best-performing neural encoder variant, when integrated into the RippleNet model, achieved NDCG@10 of 0.0004, Precision@10 of 0.0002, and Recall@10 of 0.0004. These results indicate that the encoder-based approach fails to meet the effectiveness requirements of our recommendation scenario. Consequently, we proceeded to evaluate a similarity-based solution.

The RippleNet model was configured with the following RippleNet hyperparameters: \texttt{n\_hop} = 5, \texttt{n\_memory} = 16, learning rate = 0.01, and trained for up to 50 epochs with early stopping based on NDCG@10 (patience = 5). These parameters were selected via a  hyperparameter tuning process with Recbole library, optimizing NDCG@10 metric, without similarity-based solution.

Table~\ref{tab:metric_results} presents NDCG@10, Precision@10, and Recall@10 for both models across the evaluated days. The production baseline outperforms RippleNet across all metrics and dates. RippleNet shows highly variable performance: it performs best on the training day (e.g., NDCG@10 of 0.02602), but its effectiveness deteriorates rapidly before and after training. Especially after training it is 3-4 times worse than on the day before training (e.g., NDCG@10 0.00302 vs 0.00914)

This bad peformance indicates that RippleNet is only effective on the day it is trained. One possible explanation is that the model heavily relies on direct knowledge propagation through entities seen during training. On the training day, most of the test-time articles are already present in the model’s graph structure, minimizing reliance on the similarity-based fallback mechanism. However, when tested on unseen content from adjacent days, the model struggles to generalize - exposing its dependence on memorized entities and possible limited use of similarity-based reasoning during inference.

\begin{table}
\footnotesize
\caption{Performance of production baseline and RippleNet with similarity-based solution across three evaluation days and three ranking metrics.}
\centering
\begin{tabular}{llccc|ccc|ccc}
\toprule
\multirow{2}{*}{Model} & \multirow{2}{*}{} 
& \multicolumn{3}{c|}{\rotatebox{90}{\textbf{NDCG@10}}} 
& \multicolumn{3}{c|}{\rotatebox{90}{\textbf{Precision@10}}} 
& \multicolumn{3}{c}{\rotatebox{90}{\textbf{Recall@10}}} \\
\cmidrule(lr){3-5} \cmidrule(lr){6-8} \cmidrule(lr){9-11}
 &  & \rotatebox{90}{\textbf{Train - 1}} & \rotatebox{90}{\textbf{Train}} & \rotatebox{90}{\textbf{Train + 1}}
   & \rotatebox{90}{\textbf{Train - 1}} & \rotatebox{90}{\textbf{Train}} & \rotatebox{90}{\textbf{Train + 1}}
   & \rotatebox{90}{\textbf{Train - 1}} & \rotatebox{90}{\textbf{Train}} & \rotatebox{90}{\textbf{Train + 1}}\\
\midrule
\makecell{Production \\ baseline}  &  
& \rotatebox{90}{0.01909} & \rotatebox{90}{0.02790} & \rotatebox{90}{0.02199} 
& \rotatebox{90}{0.01127} & \rotatebox{90}{0.01420} & \rotatebox{90}{0.01068 }
& \rotatebox{90}{0.03551} & \rotatebox{90}{0.04313} & \rotatebox{90}{0.03118} \\
\midrule
\makecell{RippleNet + \\ Similarity}  &  
& \rotatebox{90}{0.00914} & \rotatebox{90}{0.02602} & \rotatebox{90}{0.00302} 
& \rotatebox{90}{0.00606} & \rotatebox{90}{0.01443} & \rotatebox{90}{0.00137} 
& \rotatebox{90}{0.01683} & \rotatebox{90}{0.04080} & \rotatebox{90}{0.00455}\\
\bottomrule
\end{tabular}

\label{tab:metric_results}
\end{table}

\subsection{Online evaluation}
Based on our previous experience, offline evaluation metrics do not necessarily correlate with online performance. Therefore, we conducted an online evaluation of the model and its deployment pipeline. The test was carried out over a two-day period, during which we trained two separate similarity-based models: one tailored for mobile users and the other for desktop users, reflecting distinct engagement patterns across platforms.

During the online test, the observed results were consistent with offline metrics. We recorded engagement uplifts of -13.77\% on mobile and -16.21\% on desktop of our business metric, indicating alignment between offline and online evaluation signals.
\section{Conclusions}

In conclusion, our proposed method effectively addresses the cold-start challenge in news recommendation by integrating RippleNet with semantic embeddings derived from large language models. 
This hybrid approach significantly enhances the ability to recommend newly published content without prior interaction data, ensuring timely and relevant content delivery. 
The empirical findings suggest that leveraging semantic relationships captured by RippleNet and LLM embeddings provides a foundation for personalized recommendations in rapidly changing environments, however at the current state such setup does not reach production-grade performance.
Future research could explore further refinements to embedding integration techniques.

\bibliographystyle{ACM-Reference-Format}
\bibliography{sample-base}

\end{document}